\documentclass{emulateapj}

\usepackage{rotating}
\usepackage{apjfonts}
\shorttitle{A {\sl Swift} look at SN~2011\lowercase{fe} }
\begin{document}
\title{A {\sl Swift} look at SN~2011\lowercase{fe}:\\
The Earliest Ultraviolet Observations of a Type I\lowercase{a} Supernova
}

\author{Peter~J.~Brown\altaffilmark{1,2}
Kyle S. Dawson\altaffilmark{1}, 
Massimiliano de Pasquale\altaffilmark{3}, Caryl Gronwall\altaffilmark{4,5}, \\
Stephen Holland\altaffilmark{6},  Stefan Immler\altaffilmark{7,8,9}, 
Paul Kuin\altaffilmark{10}, Paolo Mazzali\altaffilmark{11,12}, \\
Peter Milne\altaffilmark{13}, 
Samantha Oates\altaffilmark{10}, \& Michael Siegel\altaffilmark{2}
}
\altaffiltext{1}{Department of Physics \& Astronomy, University of Utah, 115 South 1400 East \#201, 
Salt Lake City, UT 84112, USA}            
\altaffiltext{2}{Current Affiliation: George P. and Cynthia Woods Mitchell Institute for Fundamental Physics \& Astronomy, 
Texas A. \& M. University, Department of Physics and Astronomy, 
4242 TAMU, College Station, TX 77843, USA; 
pbrown@physics.tamu.edu}            

\altaffiltext{3}{Department of Physics and Astronomy, University of Nevada, Las Vegas - 4505 S. Maryland Parkway,
		Las Vegas, NV 89154, USA}
\altaffiltext{4}{Department of Astronomy and Astrophysics, 
				The Pennsylvania State University, 
				525 Davey Laboratory, 
				University Park, PA 16802, USA}
\altaffiltext{5}{Institute for Gravitation and the Cosmos, 
				The Pennsylvania State University, 
				University Park, PA 16802, USA}

\altaffiltext{6}{Space Telescope Science Center
3700 San Martin Dr., 
Baltimore, MD 21218, USA}
\altaffiltext{7}{Astrophysics Science Division, Code 660.1,
                8800 Greenbelt Road
                Goddard Space Flight Centre,
                Greenbelt, MD 20771, USA}
\altaffiltext{8}{Department of Astronomy, University of Maryland, College Park, MD 20742, USA}
\altaffiltext{9}{Center for Research and Exploration in Space Science and Technology, 
NASA Goddard Space Flight Center, Greenbelt, MD 20771, USA}
\altaffiltext{10}{Mullard Space Science Laboratory, University College London, Holmbury St. Mary, Dorking Surrey, RH5 6NT, UK}            
\altaffiltext{11}{Max-Planck-Institut fur Astrophysik, Karl-Schwarzschild-Strasse 1, D-85748 Garching, Germany}
\altaffiltext{12}{INAF-Osservatorio Astronomico, vicolo dellâOsservatorio, 5, I-35122 Padova, Italy}

\altaffiltext{13}{Steward Observatory, University of Arizona, Tucson, AZ 85719, USA}            


\begin{abstract}

We present the earliest ultraviolet (UV) observations of the bright Type Ia supernova 
SN~2011fe/PTF11kly in the nearby galaxy M101 at a distance of only 6.4 Mpc. 
It was discovered shortly after explosion by the Palomar Transient Factory and 
first observed by Swift/UVOT about a day after explosion. 
The early UV light is well-defined, with $\sim$20 data points per filter in the 
five days after explosion.  These early and well-sampled UV observations form new template 
light curves for comparison with observations of other SNe Ia at 
low and high redshift.  We report fits from semi-empirical models of the explosion and 
find the time evolution of the early UV flux to be well fit by the superposition of two 
parabolic curves.  
Finally, we use the early UV flux measurements to examine a possible shock interaction 
with a non-degenerate companion. From models predicting the measurable shock emission, we find that even a solar mass companion at a distance 
of a few solar radii is unlikely at more than 95\% confidence.

\end{abstract}

\section{Early Observations of Type I\lowercase{a} Supernovae}  \label{intro}

The first electromagnetic signal of a supernova (SN) occurs when the explosive shock 
breaks through the surface of a star or its optically thick circumstellar envelope \citep{Colgate_1974,Klein_Chevalier_1978}.
It is characterized by a rapid rise in luminosity, with a spectrum peaking at X-ray/ultraviolet (UV) wavelengths, 
that quickly fades.  This shock breakout is most often discussed in the context of core-collapse SNe models, and because it 
only lasts briefly, well before the SN becomes optically bright,    
observing it requires an external trigger or frequent monitoring.  Several such observations now exist from the past $\sim$20 years.
Observations of  SN 2006aj were triggered by the accompanying gamma ray burst (GRB) 060218 detected by the Swift spacecraft \citep{Gehrels_etal_2004, Campana_etal_2006}, and the observations are consistent with the shock breakout from a dense circumstellar wind \citep{Campana_etal_2006, Modjaz_etal_2006, Pian_etal_2006, Soderberg_etal_2006, Mirabal_etal_2006, Sollerman_etal_2006, Mazzali_etal_2006, Li_2007, Sonbas_etal_2008}.  
The shock breakouts of SNe  SNLS-4D2dc and SNLS-06D1jd \citep{Gezari_etal_2008,Schawinski_etal_2008},  and PTF 09uj \citep{Ofek_etal_2010} were serendipitously observed in the UV by GALEX \citep{Martin_etal_2005} and 2008D  \citep{Soderberg_etal_2008,Modjaz_etal_2009} by Swift's Ultra-Violet/Optical Telescope (UVOT; \citealp{Roming_etal_2005}).  
Wide field, high cadence, coordinated surveys increase the chances of discovering SNe during shock breakout and acquiring high quality data. 
Even after the shock breakout has occurred, rapid response observations in the UV can observe the cooling of the shock \citep{Kirshner_etal_1987,Fransson_etal_1987, Roming_etal_2009, Gal-Yam_etal_2011, Arcavi_etal_2011}, yielding valuable clues about the nature of the progenitor and its environment.

While not previously available, high cadence, high signal-to-noise ratio (S/N) measurements
of early SNe~Ia could similarly reveal the size of the progenitor and the nature of the explosion.  
In particular, a transition from a deflagration to a supersonic detonation 
should result in a breakout shock, observable in the first few hours at X-ray/UV energies 
\citep{Hoeflich_Schaefer_2009,Piro_etal_2010,Rabinak_etal_2011}.
Early data can also test the assumptions underlying 
the commonly used, parabolic fireball model \citep{Riess_etal_1999} and test how well the explosion date can be determined by the extrapolation of that model.

Observations in the first few days can also constrain the size and 
separation of a companion star (\citealp{Kasen_2010,Brown_etal_2012}, hereafter K10 and B12) or circumstellar material in the progenitor system \citep{Hoeflich_Schaefer_2009,Fryer_etal_2010} by comparison with the predicted luminosity.  
While SNe~Ia are important for cosmology (\citealp{Riess_etal_1998, Perlmutter_etal_1999}; see also \citealp{Weinberg_etal_2012} for the role of SNe in the context of other cosmological probes),  their progenitor systems are 
not well understood.  This is cause for concern because the progenitor systems might evolve with cosmic time leading to a systematic change in the properties of the explosion.  An evolving population of SNe~Ia progenitors could be mistaken for distinct models of dark energy \citep{Podsiadlowski_etal_2006,Riess_Livio_2006}.

While the SN~Ia progenitor is widely believed to be a degenerate Carbon-Oxygen white dwarf (WD) in a binary system, the companion could be another WD (the double degenerate scenario) or a red giant (RG) or main sequence (MS) star (the single degenerate scenario).  In the double degenerate scenario \citep{Iben_Tutukov_1984,Webbink_1984, Livio_2000}, the orbital separation between the WDs shrinks until they merge or the less massive WD is disrupted and accreted onto the SN progenitor.  In the single degenerate scenario the companion donates mass to the progenitor via Roche-lobe overflow \citep{Whelan_Iben_1973} or 
a stellar wind \citep{Hachisu_etal_1999}.  
Comparison of observations to the K10 models for the interaction of the SN~Ia ejecta with its companion allow a determination of the separation distance for the case of a Roche-lobe filling, non-degenerate companion.  
Previous work has used large samples of early optical data 
to rule out RGs as the primary companions of SNe~Ia progenitors 
\citep{Hayden_etal_2010b, Tucker_etal_2011, Bianco_etal_2011, Mo_etal_2011}.  
In B12 we used early UV observations from a sample of twelve SNe~Ia to place 
similar limits on the companion.

Here we present results from very early {\sl Swift} 
observations of SN~2011fe in the nearby galaxy M101,
the earliest UV measurements to date for a SN~Ia.  In Section \ref{obs} we describe the data reduction and present the most densely sampled set of UV observations for any SN~Ia observed to date. We present $\sim20$ data points per filter within five days after explosion and over one thousand data points in the two months after explosion.  In Section \ref{analysis} we use these measurements  to create more accurate UV templates, compare the early flux with the fireball model, and use the lack of observed shock emission predicted in the K10 models to push the constraints 
to smaller companion sizes than in B12.  The implications of this analysis are summarized and discussed in Section \ref{conclusion}.

 
\section{Observations} \label{obs}

  
SN~2011fe, also known as PTF11kly, was discovered in M101 at a magnitude g=17.2, 
classified as a probable young Ia, and promptly announced 
 by the Palomar Transient Factory (PTF; \citealp{Law_etal_2009}) 
on 2011 August 24 \citep{Nugent_etal_2011a}.  
The first PTF detection was August 24.167 \citep{Nugent_etal_2011b}.  
It was not detected by PTF to a limiting magnitude of 21.5 one day before, 
strongly constraining the explosion date estimated by \citet{Nugent_etal_2011b} 
to be August 23.687 $\pm$ 0.014 from a power law fit to the first three nights of PTF g-band data.   
X-ray and UV observations were promptly requested from the {\sl Swift} observatory, 
and observations began August 24.9.  
{\sl Swift}'s Ultraviolet/Optical Telescope (UVOT; \citealp{Roming_etal_2005}) 
utilized the 6 broadband filters with the following central wavelengths ($\lambda_c$)  
and full-width half maximum (FWHM) in Angstroms: uvw2 ($\lambda_c$=1928; FWHM=657), 
uvm2 ($\lambda_c$=2246; FWHM=498), uvw1 ($\lambda_c$=2600; FWHM=693), u ($\lambda_c$=3465; FWHM=785),
b ($\lambda_c$=4392; FWHM=975), and v ($\lambda_c$=5468; FWHM=769).  
Initial UVOT magnitudes were reported by \citet{Cenko_etal_2011b} and X-ray 
upper limits from {\sl Swift}/XRT by \citet{Margutti_Soderberg_2011}.  

Following the announcement of the discovery of SN~2011fe, 
we requested daily {\sl Swift} observations to monitor its UV and optical behavior.
A multi-filter image of SN~2011fe and its host galaxy is displayed in Figure 1.  SN~2011fe rapidly brightened, necessitating several changes to the normal 
SN observing strategy and data reduction.  
After the first several observations we changed observing modes 
to use a smaller region of the CCD 
read out at a faster rate (3.6 ms compared to the normal 11.0 ms frame time) 
so the effects of coincidence loss could be corrected 
to a higher count rate \citep{Poole_etal_2008}.  
Observations with more than 0.95 counts per frame were discarded 
due to the larger uncertainties on the coincidence loss correction 
as the source brightness approached and passed the point of saturation (see e.g. \citealp{Kuin_Rosen_2008}).  
The use of smaller hardware windows allowed us to follow SN~2011fe to 
magnitudes of 11.26, 12.44, and 10.82, in the u, b, and v filters, respectively.
However, the detectors began to saturate at count rates fainter than the peak of the light curve in 
any of the optical filters. 
In the UV, count rates are much lower, but near peak the SN still 
required significant corrections to the UV rates and some frames 
were saturated in the uvw1 filter.

The adopted analysis generally follows the procedure of \citet{Brown_etal_2009}. 
The standard UVOT aperture is 5\arcsec ~ \citep{Poole_etal_2008}, though a smaller aperture (3\arcsec~in \citealp{Brown_etal_2009}) with a corresponding aperture correction is often used to maximize the S/N.  For most of the observations a 5\arcsec~ aperture was used as the S/N was sufficiently high that the uncertainty in the aperture correction would be much larger than the photometric uncertainty.  For the fainter epochs (fainter than about 17 mag) in the UV 
the 3\arcsec~aperture was used as it gave the higher S/N.
Pre-explosion images of M101 taken in 2007 March/April (see Figure 1)
were used to subtract the underlying galaxy count rate.  
This approach is taken instead of subtracting the actual images as is usually done 
with linear CCD observations (e.g.  \citealp{Alard_Lupton_1998} ) so that the coincidence loss 
correction can be made on the observed galaxy count rates and the observed galaxy+SN count 
rates individually before the subtraction.   
The coincidence loss corrected count rates are given in Table 1 along with 
the apparent magnitudes.
The final data set 
uses over 1000 individual exposures, including $\sim$20 points per filter in the first 
five days after explosion and $\sim$50 pre-maximum points per filter in the UV. 
The photometry is based on the updated UVOT photometric system of \citet{Breeveld_etal_2011} 
and includes the time dependent sensitivity correction. The analysis below uses the updated effective 
area curves for the UVOT filters.
A Cepheid-based distance modulus of 29.04 $\pm$ 0.20 (6.4 Mpc; \citealp{Shappee_Stanek_2011}) is assumed for 
the absolute magnitudes.  A small reddening of E(B-V)=0.01 in the direction of the SN Ia is assumed for 
the Milky Way \citep{Schlegel_etal_1998} and the host galaxy 
reddening is negligible \citep{Li_etal_2011}.


\section{Analysis}\label{analysis}

The excellent sampling of this data enables a detailed look at the early 
UV behavior for the purposes of making template light curves, modeling the early rise compared to the fireball model, and putting constraints on single degenerate companions.

\subsection{Early UV light curves and colors}\label{lightcurves}

\begin{figure} 
\resizebox{16cm}{!}{\includegraphics*{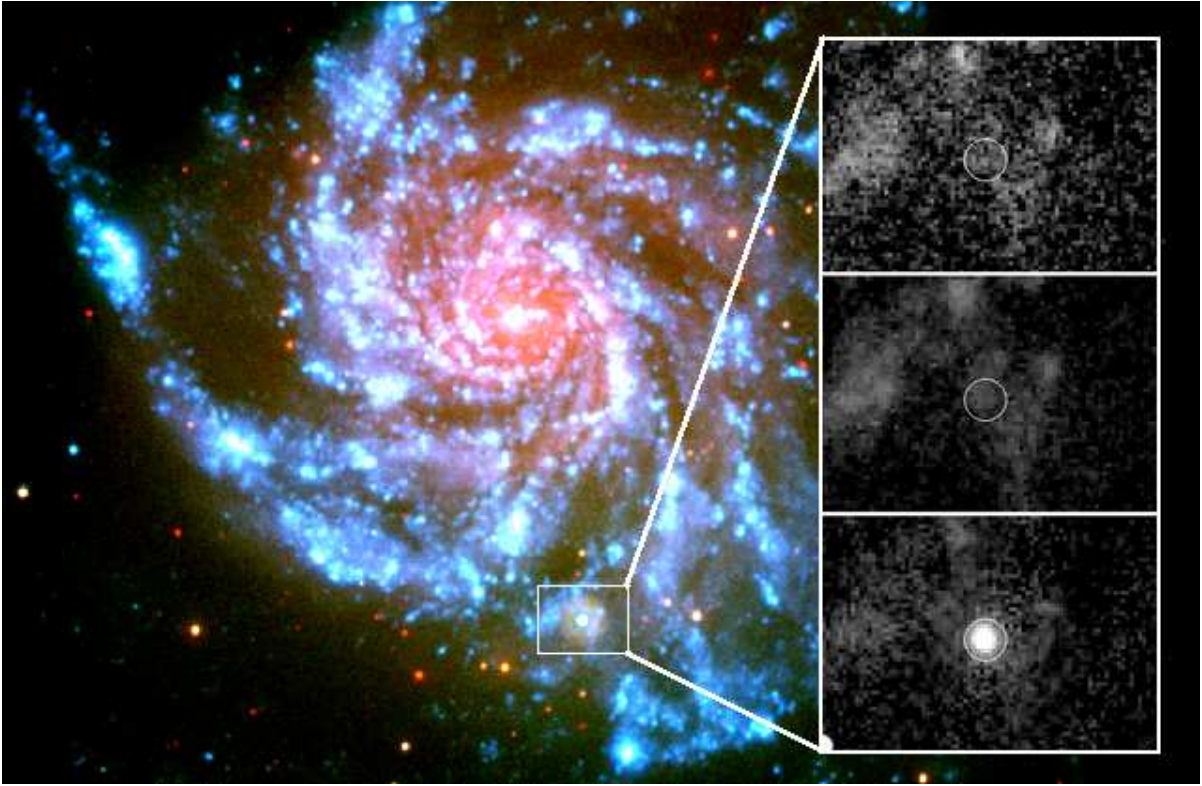}  }
\caption[Results]
        {UVOT image of M101 and SN~2011fe in the uvm2,uvw1, and v filters of UVOT.  
The inset (80 \arcsec~ by 60 \arcsec) shows uvm2 images of the area around SN~2011fe in pre-explosion images, the first observations after discovery, and near peak.  [This figure is available in color in the electronic version, with the red, green and blue channels corresponding to v, uvw1 and uvm2 respectively.]

 } \label{plot_images}    
\end{figure}

\begin{figure} 
\resizebox{16cm}{!}{\includegraphics*{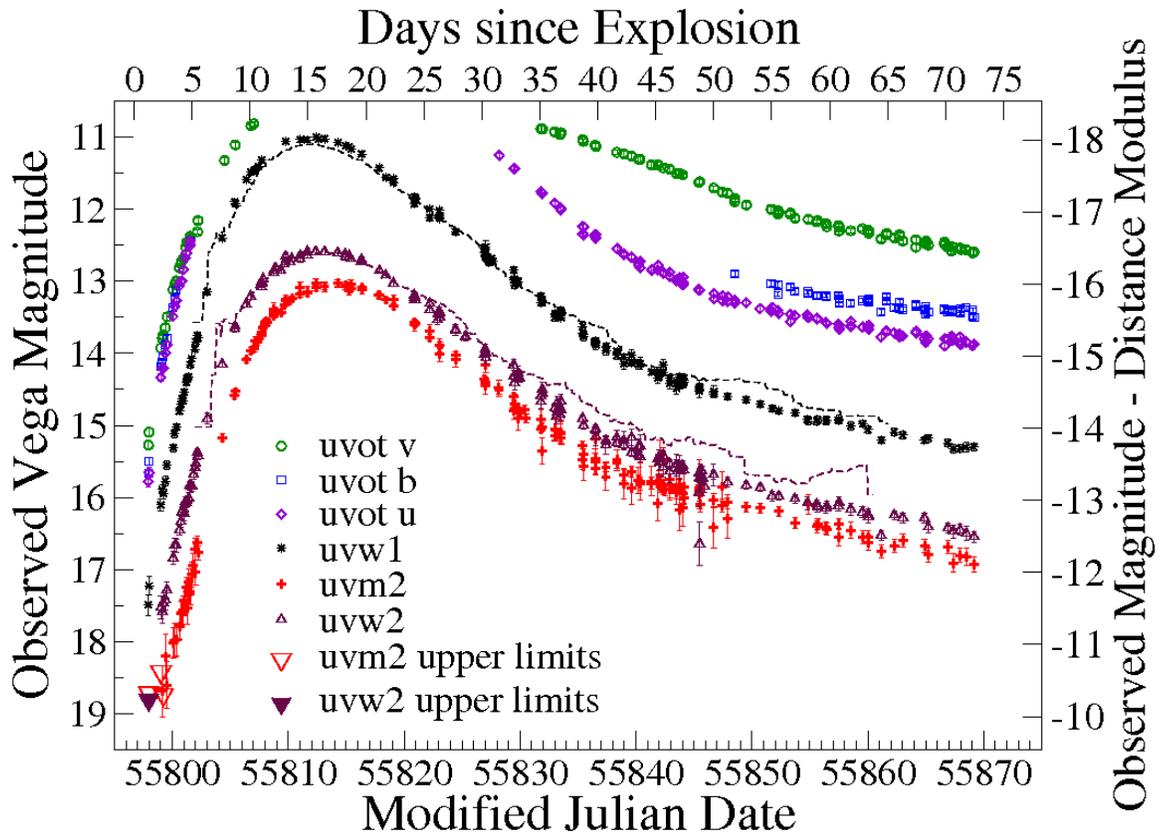}   }
\caption[Results]
        {UVOT light curves of SN~2011fe in Vega magnitudes.  
The M10 templates for uvw1 and uvw2 are overplotted with dashed lines.
 
 } \label{plot_lightcurves}    
\end{figure}

Figure \ref{plot_lightcurves} displays the exquisitely sampled UVOT light curves of SN~2011fe.  
While the SN had already brightened to $\sim$15.7 mag in the optical $\sim$1 day after explosion, 
the first two exposures in uvm2 provided only 99\% upper limits at mag 19.2 
(corresponding to an absolute magnitude of -9.6 
and a flux density of $\sim 5 \times 10^{-17}$ erg s$^{-1}$ cm$^{-2}$ \AA$^{-1}$).  
This first epoch of uvm2 is displayed in the inset of Figure \ref{plot_images}.  

SNe~Ia have long been characterized by their low UV flux relative to the optical 
at maximum light \citep{Holm_etal_1974,Kirshner_etal_1993,Panagia_2003}. 
Early observations of SNe~Ia reveal an even larger deficit of UV flux (\citealp{Milne_etal_2010}, hereafter M10).  
The very early observations of SN~2011fe allow us to examine the behavior right after the explosion. 
 Figure \ref{plot_colors} shows the uvm2-uvw1 and uvw1-v color evolution 
of SN~2011fe.  
In the first few days after explosion, the colors are very red (i.e. fainter at shorter wavelengths) and nearly constant before becoming bluer 
with time like other SNe~Ia observed with Swift (M10).  
SN~2009ig, whose UVOT observations began about two days before explosion did not show this plateau but was becoming bluer already at the onset of observations \citep{Foley_etal_2011}.  The colors of normal SNe~Ia reach a minimum a few days before 
optical maximum light.  The early UV deficit is believed to be caused by a 
lack of heavy elements in the outermost layers ($>12-15000$ km s$^{-1}$) of the 
SNe at early times.  In this scenario, UV photons will be absorbed at smaller radii, 
and the outer layers do not have the composition to produce inverse fluorescence \citep{Mazzali_2000}. 
As the SN photosphere recedes with time, UV photons will still be absorbed, 
but larger abundances of Fe, Co, Cr, Ti will be present near the photosphere.  
The optical lines of FeII, III, Co II, III, Ti II, CrII are expected to saturate, 
and fluorescence via UV lines should then become possible.  
As the SN approaches maximum optical light, a decrease in temperature leads again 
to a reddening in the uvw1-v color.
The color evolution of SN~2011fe is shifted blueward from the average SNe~Ia (M10).  
Combined with the detection of CII in the 
early spectra \citep{Cenko_etal_2011a}, this is consistent with the observation that SNe~Ia with 
carbon usually have bluer NUV-optical color evolution 
\citep{Thomas_etal_2011, Milne_Brown_2012}.

It is essential to model the time evolution of SN~Ia luminosity through 
template light curves to determine times of maximum light, 
interpolate light curves, differentiate 
between typical and atypical SNe, and define normal behavior for comparison 
with theoretical models.  The first near-UV SN~Ia template 
(F275W filter with peak wavelength = 2740 \AA ~  and FWHM=594 \AA)  was 
generated from International Ultraviolet Explorer (IUE) and Hubble Space Telescope (HST) 
observations of SNe 1990N and 1992A \citep{Kirshner_etal_1993} .  
This served as an excellent 
template for early {\sl Swift}/UVOT observations \citep{Brown_etal_2005} 
without the stretching usually required in the optical to fit individual SNe.  
M10 improved upon this template using normal events observed by {\sl Swift}/UVOT.  
Only the rapid declining SNe 2005ke \citep{Immler_etal_2006} and 2007on 
(which were not included in the generation of the template) 
show significant deviations from it (M10).   

The early, frequent, and high S/N observations of SN~2011fe make it an 
excellent template for comparison with other SNe Ia.  It is generally consistent 
with the average template from M10 and has about the same number of data points 
as the whole set of SNe used in its construction, but it avoids some of the 
complications of combining unevenly sampled data points from objects 
which may or may not have similar light curve shapes.  
In particular, SN Ia light curves in the uvm2 filter (\citealp{Brown_etal_2009}, M10) 
exhibit too much variety to create an average or 
composite template.  To create a smooth, uniformly sampled template, we fit the rise, peak, and 
decay of SN~2011fe's UV light curves with high order polynomials.  These are spliced together where they 
overlap and given in Table 2.  
We note that the previous earliest UV observations
from SN~2009ig (\citealp{Foley_etal_2011};B12), 
can be stretched (i.e. scaling the time axis) to match the SN~2011fe templates.
The stretching must be done independently before and after maximum 
as in \citet{Hayden_etal_2010a}, as SN~2009ig rises more quickly but then fades more slowly.
While the UV light curves of SNe Ia are more similar in shape than their optical light curves (M10), 
differences are noticeable for the SNe with extremely broad or narrow optical light curves.  The increasing number of early and well sampled UV light curves should yield valuable insights into 
their true diversity and any correlations with the optical or UV brightness.

The time and magnitude at maximum brightness has been found in each filter by finding where 
the derivative of the polynomial fit equals  zero.  The peak magnitudes for the uvw2, uvm2, and uvw1 filters are 12.59, 13.06, and 11.02, respectively.  The peak times (in MJD) for the uvw2, uvm2, and uvw1 filters are 55813.0, 55813.4, and 55812.4, respectively.  The peak magnitudes for uvw2 and uvw1 are consistent with that determined by matching up the M10 templates using 
$\chi^2$ minimization of the differences.
Subtracting the distance modulus of 29.04 $\pm$ 0.20 gives absolute magnitudes of -16.45, -15.98, and -18.02, comparable to other SNe~Ia observed in the UV \citep{Brown_etal_2010}.  

\subsection{The Expanding Fireball Model and the Early UV Flux} \label{fireball}

The early optical flux curves of SNe~Ia are often assumed to follow the ``expanding fireball'' 
model described in \citet{Riess_etal_1999}.  
Assuming that the flux arises from a quasi-blackbody observed on the Rayleigh-Jeans tail, 
the expanding photosphere would have an emitting area proportional to 
the square of the velocity and the square of the time since explosion squared.  
If the temperature and velocity are relatively constant compared to the rapidly changing time 
since explosion, 
then those other terms can be assumed into a constant of proportionality.  
Specifically, the flux relates to the time since explosion approximately as $f=\alpha (t-t_0)^2 $ \citep{Riess_etal_1999, Garg_etal_2007, 
Mo_etal_2011}, where t is the observation date, t$_0$ is usually taken to be the date of explosion, 
and $\alpha$ is a constant that absorbs the distance, temperature, velocity, and other factors.  
The flux is zero for $t<t_0$.
The assumptions underlying the use of the fireball model in the optical are not as applicable in the UV.   
UV SN flux does not come from the Rayleigh-Jeans tail of a blackbody spectrum--
the little flux emitted from the thermal photosphere is mostly absorbed by a dense forest of 
absorption lines from iron-peak elements \citep{Pauldrach_etal_1996} and most of the UV light 
which is observed results from reverse fluorescence \citep{Mazzali_2000}. 
We will nevertheless use the fireball model as a starting point for comparisons.

The conversion from observed count rate to flux requires a spectrum-dependent conversion factor for the Swift bandpass filters \citep{Poole_etal_2008}. To estimate this factor for each epoch of photometry, we have have taken the closest epoch spectrum from a SN~Ia spectral series \citep{Hsiao_etal_2007}  and warped it to match the observed count rates (excluding uvw2 as its effective wavelength is very spectrum dependent) using a 2nd order polynomial and three iterations of warping. At the epochs where the SN~2011fe optical data were saturated, we interpolated from the observed UVOT count rates of a similar SN~Ia (SN~2005cf) scaled to match the pre and post-peak data of SN~2011fe.  To test the sensitivity of the results to the input spectrum, we also performed the analysis using the HST spectrum of SN~1992A \citep{Kirshner_etal_1993}, a 6000 K blackbody spectrum, and a flat (constant flux density versus wavelength) spectrum. We note that we calculated conversions between the observed count rate and the integrated flux, and these are less sensitive to the details of the spectrum than the flux density factors calculated by \citet{Brown_etal_2010}. Nevertheless, the different spectra change the conversion factors by less than 5\% in the optical filters, 9\% in the uvw1 filter, and 6\% in the uvm2 filter.  The variation is as large as 15\% in the uvw2 filter due to its larger wavelength range and the difficulty in constraining the spectral warping at the short wavelength end.  The change in the factors with time also differ between the models, especially in the UV.  While the most accurate modeling would require the UV spectra or at least a more similar template, the features noted below are qualitively similar regardless of the template spectrum used and are also visible in the uncorrected count rate curves.  
The integrated flux in each filter at each epoch is given in Table 1. 
We wish to emphasize that the best comparison with theoretical models would not be with the model-dependent fluxes but by computing spectrophotometry on the models themselves and comparing them with the observed magnitudes or count rates.

Figure \ref{plot_fireball} shows the flux curves over the first ten days after explosion 
along with the best fit parabolic curves.  
The fitting was performed with the routine MPFITFUN.pro which utilizes the Levenberg-Marquardt Algorithm \citep{Markwardt_2008,More_1978}.
The fit parameters are given in Table 3 for different epoch ranges of the data. 
The UVOT b and v curves can be fit with explosion dates of August 23.79 and 23.62, respectively, bracketing the explosion date of August 23.687 calculated by \citet{Nugent_etal_2011b} from g-band data.  All of our pre-maximum optical data is consistent with the fireball model, though the data set is limited in time by the saturation issues.

The UV fits for the first four days are also consistent with the fireball model. 
As the UV fits are expanded beyond five days after the explosion, the quality of the fits are drastically reduced, 
as the count rate rises quicker than the extrapolated model.  
For example, fitting the uvm2 count rates for the exposures less than four days after explosion, 
a t$_0$ of August 23.81 $\pm$ 0.28  is found, consistent with the optical filters. If data between 5 and 10 days 
after explosion are used (more typical for early observations of SNe~Ia), 
a larger amplitude is found and a much later t$_0$ of August 26.76 $\pm$ 0.30, 
which clearly does not correspond to the explosion date fit by the earlier data. 
The uvw1 and uvw2 flux exhibit similar behavior. The optical tails of the uvw2 and uvw1 filters would only dilute this feature seen in the UV (in particular the uvm2 filter which has no significant 'red leak') and not the optical filters. Extrapolating the parabola fit to the 5-10 day observations back to the time of the earlier 
observations, the observed early flux would appear as an excess compared to the fireball model. 
Excess UV flux in the earliest observations compared to a fireball model was also found by \citet{Foley_etal_2011} 
in SN~2009ig but rejected as evidence of shock interaction with a companion because of 
the color evolution.

To address the apparent change in the early slope, 
we introduce a second component to the fireball model:

$f= \alpha_1 (t-t_{0,1})^2 +\alpha_2 (t-t_{0,2})^2$


These best fit parameters are given in Table 3 
for the three UV filters and the v filter (the only optical filter with unsaturated data covering the epochs of interest).  The reduction in the reduced  $\chi^2$ compared 
to a single parabolic fit over the same ten day range is dramatic in the UV but insignificant in the v band.
In an attempt to simulate a possible shock breakout, we also tried a second model 
consisting of an early bump parameterized as a parabola with a negative amplitude 
superimposed on a fireball model.  However, the fit gave a $\chi^2$ nearly triple that 
of the double fireball model and was rejected.  As discussed by \citet{Foley_etal_2011} 
for SN~2009ig, the reddening of the colors is also inconsistent with a cooling shock.


\subsection{The unseen shock from a companion}\label{noshock}

The early time UV data from SN~2011fe is also important for what is not seen 
-- excess UV emission arising from the interaction between the SN explosion 
and the companion (K10).  
In the single degenerate Roche-lobe overflow scenario, 
this interaction is predicted to produce a shock that is very bright 
in the first few days after the explosion, particularly in the UV. 
In B12, we used numerical and analytic models from K10 to predict the luminosity 
of this shock as a function of viewing angle and companion separation distance.  
The analytic models give the time dependent luminosity and temperature as a function 
of the separation distance.  From these we calculate the expected brightness 
of the shock in the 6 UVOT filters.  
The peak luminosity of the shock emission increases for 
larger separation distances 
(and thus larger stellar radii of the companion, since it is assumed to fill its Roche-lobe).  
Thus, a 1 M$_\sun$ evolved red giant (RG) companion at a separation distance of 
2 $\times 10^{13}$ cm produces more UV shock emission than main sequence (MS) stars. 
For all companions, the maximum shock emission 
occurs for a viewing angle of 0 degrees, corresponding to a geometry in which 
the companion lies directly in the line of sight between the observer and the SN~Ia.  

Following the method of B12, we do not attribute any observed UV flux to the SN~Ia, 
but instead use it as an upper limit on the early UV flux from the shock. 
This is necessary because the independent UV templates of M10 
do not begin as early as these 
observations and because numerical simulations do not adequately match the 
observed UV light of SNe~Ia (B12).    
Spectrophotometry from the modeled spectra are compared to the observations as in B12, including the optical tails of the uvw2 and uvw1 filters (often referred to as the `red leaks'). 
We improve the analysis of B12 for the fainter observations by comparing predicted and observed count rates rather than magnitudes.  We determine 95\% confidence lower limits on the viewing angle 
for each separation distance through 
Monte Carlo realizations that model the errors in the explosion date, observed count rates, 
distance modulus, and reddening.  
Further details of the analysis are found in B12.

For SN~2011fe, the very early and deep UV observations result in tighter limits 
on the shock luminosity than any SN~Ia in B12.  As with most of the 
SNe~Ia in that sample, the strictest limits come from the first observations 
in the uvm2 filter.  In the SN~2011fe data, the 95\% upper limit on the 
absolute magnitude is uvm2$>$-9.6 mag ($\sim 5 \times 10^{-17}$ ergs s$^{-1}$ cm$^{-2}$ \AA$^{-1}$)  
at 1.2 days after the estimated time of explosion (August 23.7 $ \pm $ 0.1). 
The left panel of Figure \ref{plot_noshock} compares the observed uvm2 count rates of SN~2011fe 
to that predicted for a 1M$_\sun$ companion at the distance of M101 for different viewing angles.
The right panel of Figure \ref{plot_noshock} compares the observed uvm2 count rates of SN~2011fe 
to that predicted for various separation distances at the distance of M101 for a viewing angle of 135 degrees.  From geometric predictions we would expect 90\% of observations to occur at angles less than this, resulting in a brighter, more easily observable shock.

Lower limits on the viewing angle are determined for a range of separation distances.
As shown in Figure \ref{plot_atheta}, the resulting lower limits on the 
viewing angle are 176 and 178 degrees for the 0.2 $\times 10^{13}$ (6 M$_\sun$ MS) and 
2 $\times 10^{13}$  cm (1 M$_\sun$ RG) separation distance models considered in B12.  
By simple geometric arguments, the probability of the SNe~Ia occurring 
at those viewing angles is negligible.  For even smaller companions, 
we obtain lower limits of 171 and 166 degrees for companions separated by 
0.05 $\times 10^{13}$ (2 M$_\sun$ MS) and 0.03  $\times 10^{13}$  cm (1 M$_\sun$ MS), 
with geometric probabilities of less than 1\% for both.   

\section{Summary} \label{conclusion}

\begin{figure} 
\resizebox{16cm}{!}{\rotatebox{90}{\includegraphics*{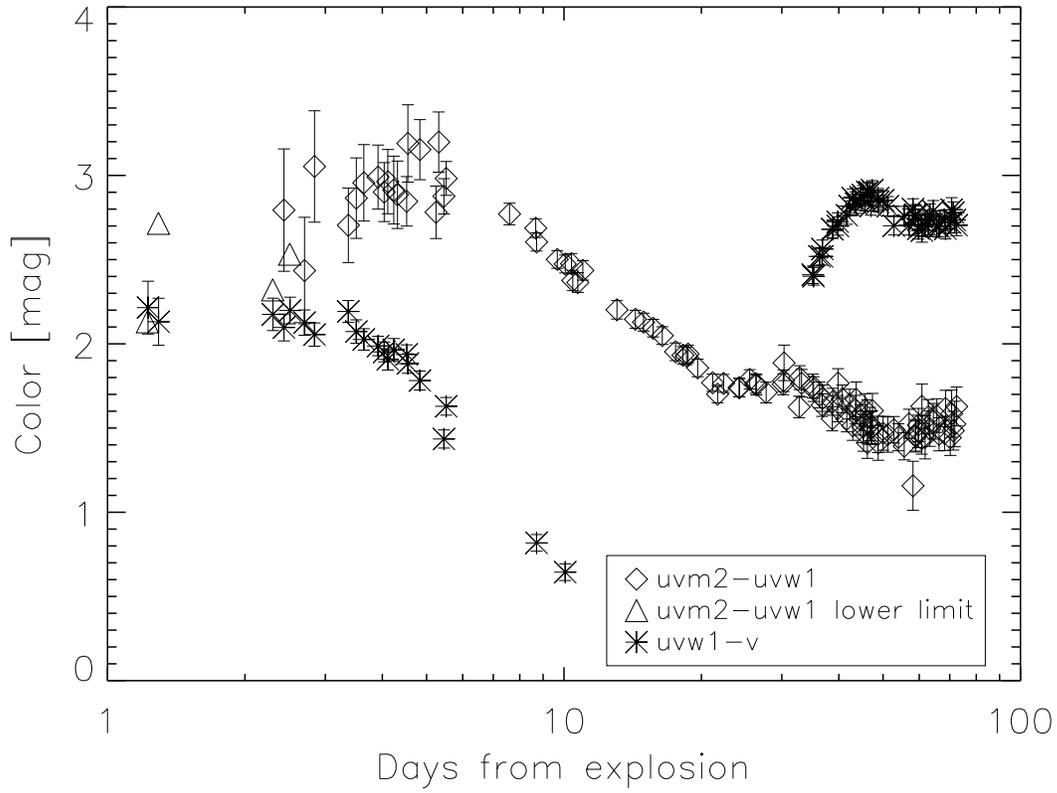}  } }
\caption[Results]
        {  Evolution in the uvm2-uvw1 and uvw1-v colors of SN~2011fe.  Errors and upper limits are one sigma.  The x-axis is plotted in log scale from the day of explosion in order to focus on the early color evolution.  The colors are constant in the first days after explosion and then get bluer.

 } \label{plot_colors}    
\end{figure}

\begin{figure} 
\resizebox{10cm}{!}{\rotatebox{90}{\includegraphics*{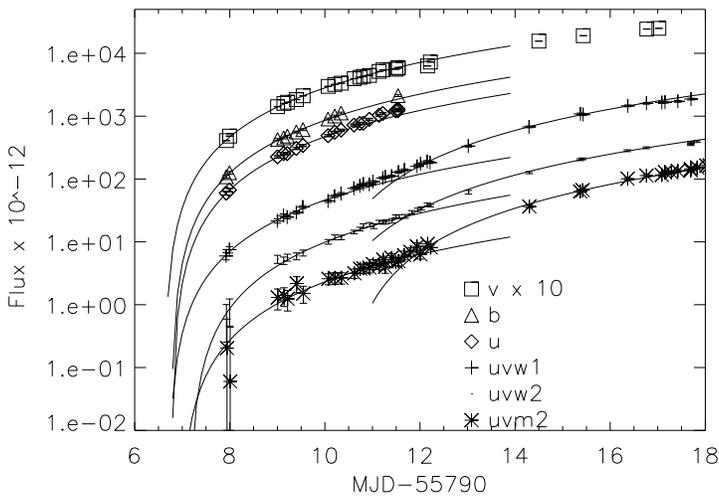}  } }
\caption[Results]
{  Integrated flux curves for five of the UVOT filters.  Fits to the early data (less than 4 days after explosion) are shown for all.  Fits to the later pre-peak data (5-10 days after explosion) are shown for the uvw1, uvm2 and uvw2 filters.   While the v data is adequately fit by a single fireball model, the UV data requires two separate components. 
} \label{plot_fireball}
\end{figure}

\begin{figure} 
\plottwo{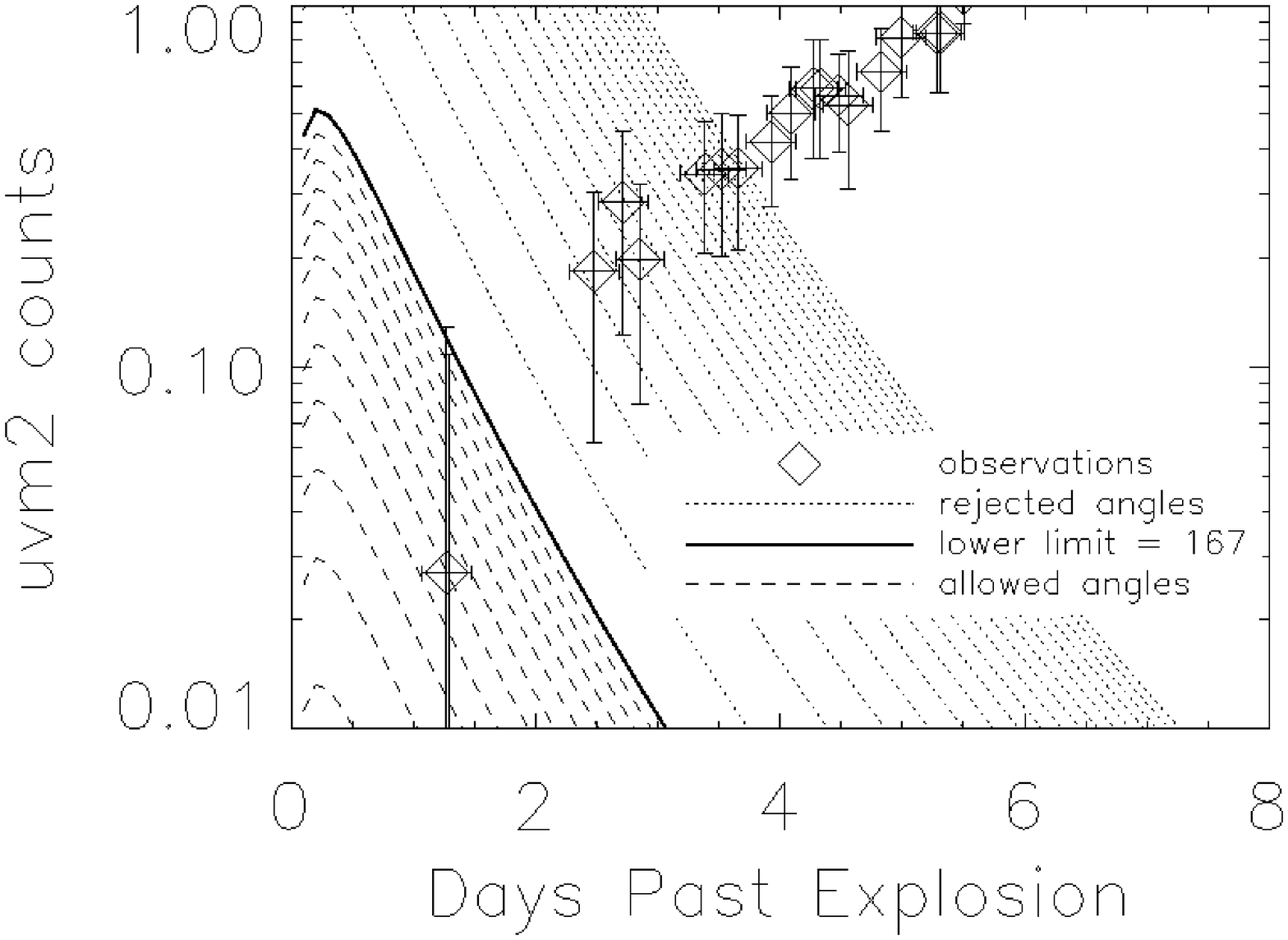}{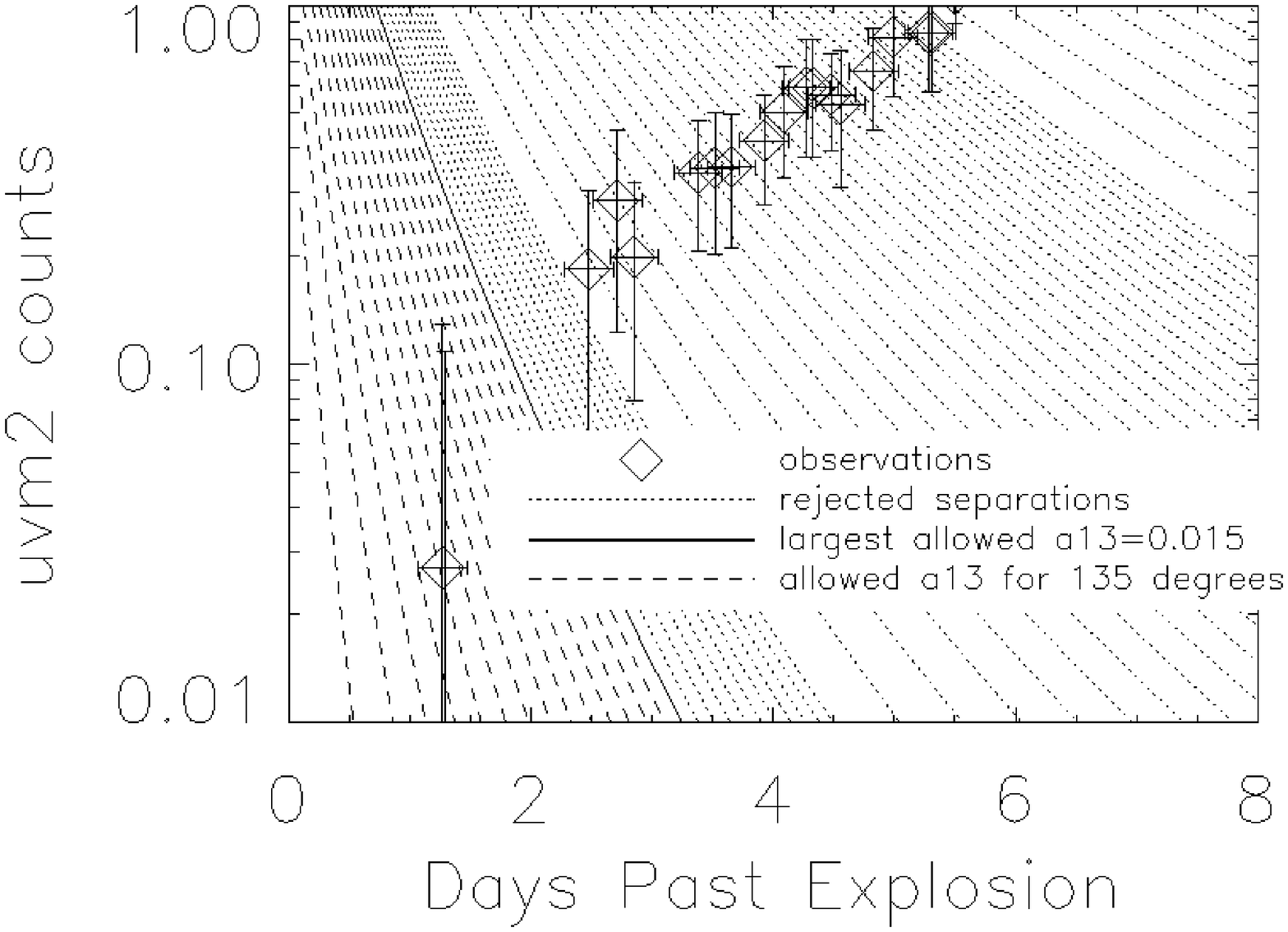}
\caption[Results]
        {{\it Left:}  Observed uvm2 count rates (with 95\% errors on the luminosity (from measured count rate, distance, and extinction) 
and 0.2 day uncertainty on the explosion date) from the first 5 exposures 
compared to the predicted count rates (K10,B12) for the 2 M$_\sun$ MS companion 
at a separation distance of 5 $\times 10^{11}$ cm for various viewing angles.  
Viewing angles at greater than 172 degrees are allowed (shown as dashed lines separated by 
one degree intervals), while those with 
smaller angles (from 0 to 170 degrees separated 10 degrees) are rejected at 95\% confidence. 
The rejected angles conflict with the first observation, and one can see 
that for this separation distance smaller viewing angles 
(and similarly for a fixed viewing angle larger separations) would have been allowed if the 
observations had not begun so soon.  
{\it Right:}  Observed uvm2 count rates (with 95\% errors on the luminosity (from measured count rate, distance, and extinction) 
and 0.2 day uncertainty on the explosion date) from the first 5 exposures 
compared to the predicted count rates (K10,B12) for a series of companion separation distances 
at a viewing angle of 135 degrees.  
The rejected models conflict with the first observation, and one can see 
that for this viewing angle much larger separation distances would have been allowed if the 
observations had not begun so soon.  
} \label{plot_noshock}
\end{figure}

\begin{figure} 
\plottwo{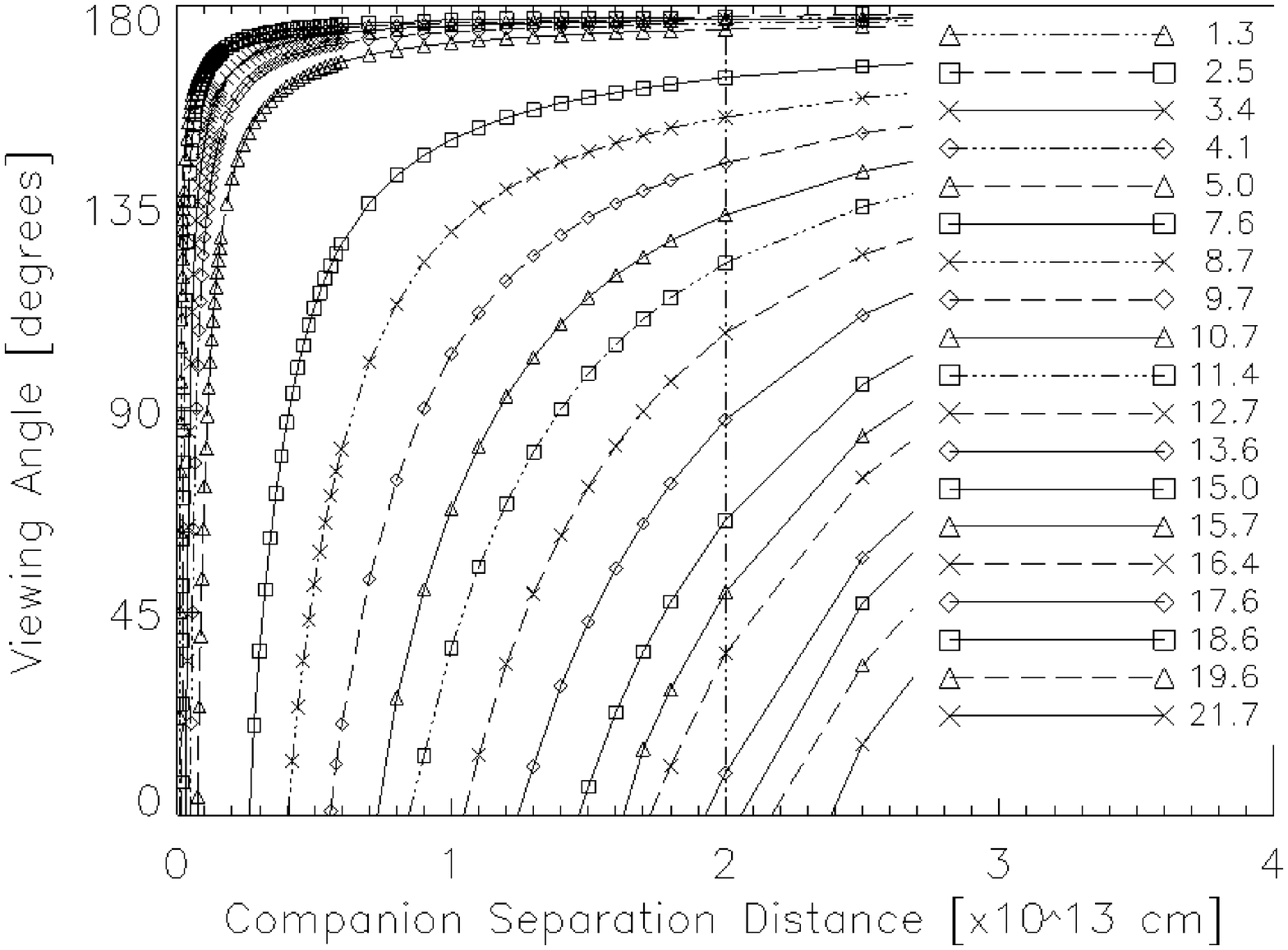}{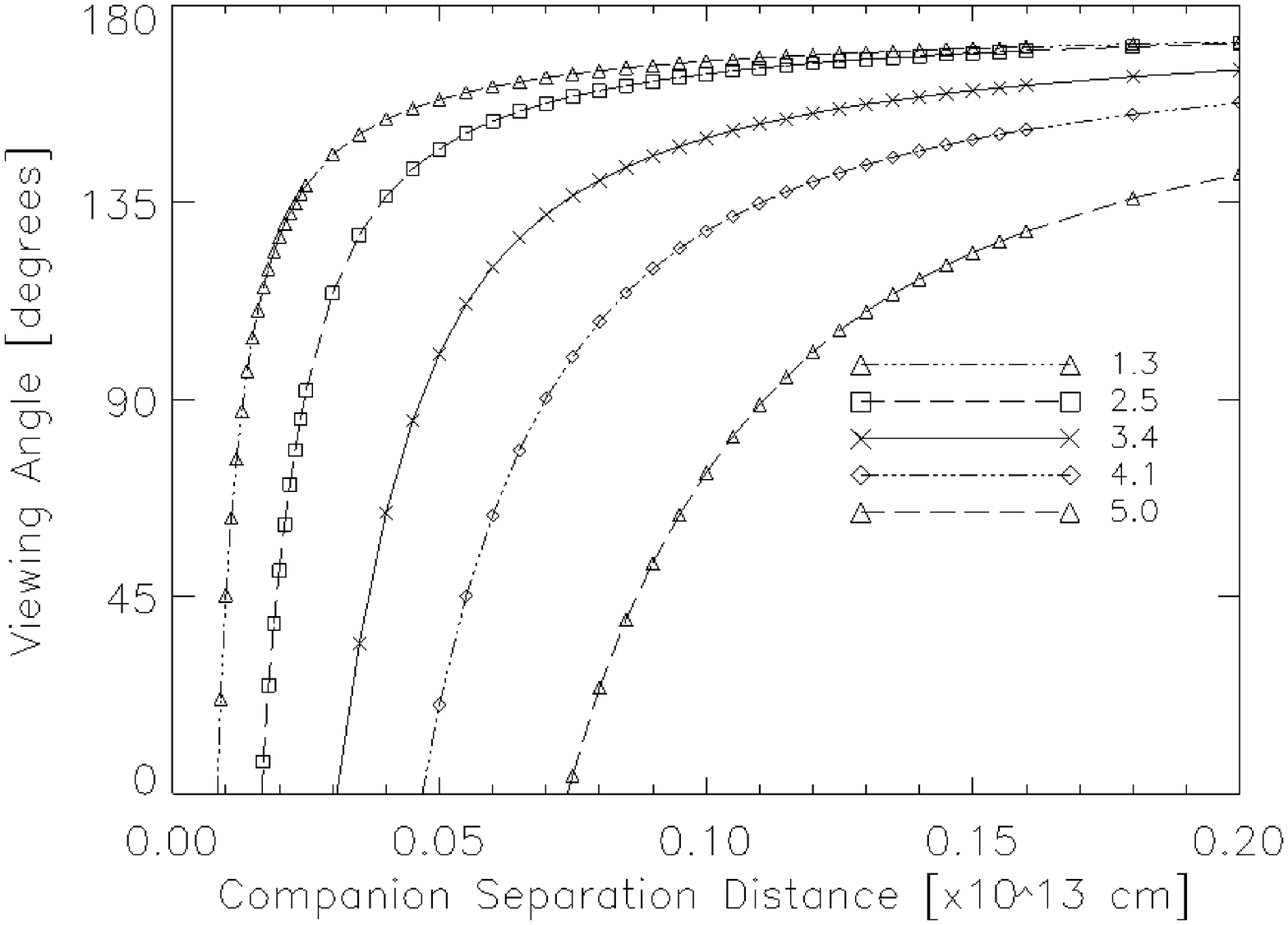} 
\caption[Results]
        {{\it Left:}  Separation distance-viewing angle constraints for SN~2011fe 
from the uvm2 filter for different epochs (given in the legend in days past explosion).  
The regions under the curve are excluded at 95\% confidence by that particular 
observation.

{\it Right:} Separation distance-viewing angle constraints for SN~2011fe 
from the uvm2 filter for different epochs (given in the legend in days past explosion).  
The regions under the curve are excluded at 95\% confidence by that particular 
observation.

} \label{plot_atheta}
\end{figure}

The early detection of SN~2011fe at such a close distance and the rapid 
response of {\sl Swift} resulted in extremely early, sensitive, and densely sampled UV measurements.  
They show the early UV/optical flux ratio to be smallest at the earliest times, but constant for the first few days after explosion, and to increase as the 
SN brightness increases.  We use the SN~2011fe to create UV light curve templates beginning one day after explosion, and comparisons with these dense and high S/N light curves will allow differences between individual SNe to be better understood.
The early flux in the optical and UV seems to follow a parabolic rise as suggested by the 
fireball model, though separate rises can be fit to the UV during the first four days 
and the period five to ten days after explosion.  The later, stronger rise might be the onset of reverse fluorescence when 
the photosphere recedes to layers inhabited by iron peak elements.  
It also coincides in time with the changing UV and UV-optical colors shown in Figure \ref{plot_colors}.  \citet{Hayden_etal_2010a} and \citet{Mo_etal_2011} point to  color evolution as a concern for the fireball model, and we show that the UV color evolution is even more problematic.  The distinct parabolic fits 
mean that data from the UV cannot be used to accurately determine the explosion date unless the 
observations begin within 5 days after explosion.

The low UV flux one day following the explosion allows us to put very tight constraints 
on the existence of a single degenerate companion in Roche-lobe overflow.  
While most previous observations could only exclude separation distances 
corresponding to RG companions (\citealp{Hayden_etal_2010b, Tucker_etal_2011, Bianco_etal_2011, Mo_etal_2011};B12), 
the limits from SN~2011fe constraining separation distances down to a few 
solar radii.  
Thus MS companions with a mass greater than 2-3.5 M$_{\sun}$, 
corresponding to the super-soft x-ray sources \citep{Li_vandenheuvel_1997,Podsiadlowski_2010}, 
are extremely unlikely.  
Very early optical observations of SN~2011fe also rule out 
RG and MS companions \citep{Bloom_etal_2012}.  Other recently published results 
further narrow down the permitted companion/accretion scenarios.  
Pre-explosion imaging from HST rules out luminous RGs and most helium stars as the 
companion \citep{Li_etal_2011}.  Limits on the X-ray luminosity \citep{Horesh_etal_2012,Chomiuk_etal_2012,Margutti_etal_2012} rule out a symbiotic RG companion donating material via stellar winds.  \citet{Nugent_etal_2011b} rule out RG and on-axis MS companions based on
the faint, early UV/optical luminosity as well as double degenerage mergers with a
dense circumstellar medium from the disrupted secondary WD.

Rather than ruling out all conventional potential progenitor systems, 
these observations do restrict the SN~2011fe system to specific conditions that 
may or may not be required for most SNe~Ia.  The companion could still be a RG or MS star if it exhausted its envelope and contracted prior to the explosion \citep{Justham_2012}.  This could happen if the accreted angular momentum prevents a prompt collapse and explosion when the SN progenitor reaches the Chandrasekhar limit.  In such a scenario, the amount of stripped Hydrogen contaminating 
spectra could be beneath observed limits \citep{Leonard_2007}.  The cross-section 
of the companion could also be small enough that its interaction with the SN ejecta (as modeled by K10 for companions still filling the Roche-lobe limit) would be much fainter than even these limits.  
\citet{Nugent_etal_2011b} ruled out WD-WD mergers because of a lack of emission from the material from the disrupted companion.   If the total mass of the system is close 
to the Chandrasekhar mass, however, most of the mass will have to be accreted before the explosion of the SN \citep{Fryer_etal_2010}.  This cleaner circumstellar environment would not result in the shocks excluded by \citet{Nugent_etal_2011b}.  
Further modeling is needed to constrain these various scenarios.  Whether these conditions are required for most SN~Ia systems will require larger samples of early observations.

These early data are a great test for the theoretical models of the early 
SN explosion itself.  The time and magnitudes reached are comparable to some models for the 
shock heated, expanded envelope of the WD itself \citep{Piro_etal_2010}, 
though \citet{Rabinak_etal_2011} predict the luminosity to be fainter by an order of magnitude 
and strongly suppressed at times greater than one hour after the explosion.  
A more detailed understanding of the early UV light is needed to 
disentangle different effects that may have been observed for the first time.
Combining these data with observations across the electromagnetic spectrum 
\citep{Nugent_etal_2011b, Horesh_etal_2012, Marion_2011, Smith_etal_2011} 
will make SN~2011fe the best studied SN~Ia ever.

\acknowledgements
We are especially grateful to the Palomar Transient Factory for promptly announcing this 
exciting object and to Eran Ofek for initating the first {\sl Swift} observations.  
This work at the University of Utah is supported by NASA grant NNX10AK43G, 
through the {\sl Swift} Guest Investigator Program.
This work is sponsored at PSU by NASA contract NAS5-00136.  
The Institute for Gravitation and the Cosmos is supported by the 
Eberly College of Science and the Office of the Senior Vice President for Research 
at the Pennsylvania State University.  
SRO and NPK gratefully acknowledge the support of the UK Space
Agency.  
This analysis was made possible by access to the 
public data in the {\sl Swift} data archive and the NASA/IPAC Extragalactic Database (NED). 
NED is operated by the Jet Propulsion Laboratory, California Institute of Technology, 
under contract with the National Aeronautics and Space Administration.  




\clearpage 

%
\begin{deluxetable}{cccccc}
\tablecaption{SN2011fe UVOT Magnitudes, Count Rates and Fluxes}\label{table_photometry}
\tablehead{\colhead{Filter} & \colhead{MJD}    & \colhead{Mag}    & \colhead{3 $\sigma$ Upper Limit}   & \colhead{Count Rate}   &  \colhead{Flux}  \\ 
            \colhead{}       & \colhead{(days)} & \colhead{(mag)} & \colhead{(mag)} & \colhead{(c s$^{-1}$)} & \colhead{(erg s$^{-1}$ cm$^{-2}$)}    } 

\startdata
uvw2 & 55797.9285 & \nodata & 18.80 &   0.10 $ \pm $  0.06 &    6.03e-13 $ \pm $    3.59e-13 \\ 
uvw2 & 55797.9954 &  \nodata & 18.80 &   0.14 $ \pm $  0.07 &    7.92e-13 $ \pm $    3.65e-13 \\ 
uvw2 & 55799.0045 & 17.53 $ \pm $ 0.16 & 18.42 &   0.87 $ \pm $  0.13 &    5.15e-12 $ \pm $    7.58e-13 \\ 
uvw2 & 55799.1339 & 17.59 $ \pm $ 0.15 & 18.56 &   0.82 $ \pm $  0.11 &    4.90e-12 $ \pm $    6.67e-13 \\ 
uvw2 & 55799.2115 & 17.46 $ \pm $ 0.14 & 18.50 &   0.93 $ \pm $  0.12 &    5.43e-12 $ \pm $    6.97e-13 \\ 
uvw2 & 55799.4016 & 17.42 $ \pm $ 0.14 & 18.48 &   0.96 $ \pm $  0.12 &    5.83e-12 $ \pm $    7.34e-13 \\ 
uvw2 & 55799.5407 & 17.29 $ \pm $ 0.12 & 18.51 &   1.09 $ \pm $  0.12 &    6.42e-12 $ \pm $    6.94e-13 \\ 
uvm2 & 55797.9407 &  \nodata & 18.70 &   0.03 $ \pm $  0.05 &    2.07e-13 $ \pm $    3.91e-13 \\ 
uvm2 & 55798.0082 &  \nodata  & 18.72 &   0.01 $ \pm $  0.05 &    5.90e-14 $ \pm $    3.69e-13 \\ 
uvm2 & 55799.0079 &  \nodata  & 18.42 &   0.17 $ \pm $  0.06 &    1.34e-12 $ \pm $    4.92e-13 \\ 
uvm2 & 55799.1391 & 18.69 $ \pm $ 0.36 & 18.70 &   0.18 $ \pm $  0.06 &    1.44e-12 $ \pm $    4.78e-13 \\ 
uvm2 & 55799.2167 &  \nodata  & 18.73 &   0.16 $ \pm $  0.06 &    1.28e-12 $ \pm $    4.60e-13 \\ 
uvm2 & 55799.4050 & 18.21 $ \pm $ 0.31 & 18.38 &   0.29 $ \pm $  0.08 &    2.27e-12 $ \pm $    6.52e-13 \\ 
uvm2 & 55799.5472 & 18.60 $ \pm $ 0.33 & 18.71 &   0.20 $ \pm $  0.06 &    1.55e-12 $ \pm $    4.68e-13 \\ 
uvw1 & 55797.9237 & 17.49 $ \pm $ 0.14 & 18.49 &   0.96 $ \pm $  0.13 &    5.61e-12 $ \pm $    7.45e-13 \\ 
uvw1 & 55797.9906 & 17.23 $ \pm $ 0.13 & 18.36 &   1.22 $ \pm $  0.14 &    7.02e-12 $ \pm $    8.24e-13 \\ 
uvw1 & 55799.0019 & 16.10 $ \pm $ 0.08 & 17.69 &   3.43 $ \pm $  0.27 &    2.03e-11 $ \pm $    1.57e-12 \\ 
uvw1 & 55799.1297 & 15.89 $ \pm $ 0.07 & 17.73 &   4.16 $ \pm $  0.26 &    2.47e-11 $ \pm $    1.52e-12 \\ 
uvw1 & 55799.2073 & 15.95 $ \pm $ 0.07 & 17.78 &   3.96 $ \pm $  0.24 &    2.32e-11 $ \pm $    1.44e-12 \\ 
uvw1 & 55799.3974 & 15.77 $ \pm $ 0.06 & 17.66 &   4.65 $ \pm $  0.27 &    2.77e-11 $ \pm $    1.61e-12 \\ 
uvw1 & 55799.5356 & 15.55 $ \pm $ 0.06 & 17.58 &   5.71 $ \pm $  0.29 &    3.35e-11 $ \pm $    1.72e-12 \\ 
u & 55797.9252 & 15.78 $ \pm $ 0.07 & 17.57 &  10.58 $ \pm $  0.68 &    5.92e-11 $ \pm $    3.79e-12 \\ 
u & 55797.9921 & 15.65 $ \pm $ 0.07 & 17.51 &  11.87 $ \pm $  0.71 &    6.63e-11 $ \pm $    3.99e-12 \\ 
u & 55799.0028 & 14.33 $ \pm $ 0.05 & 16.55 &  40.07 $ \pm $  1.74 &    2.24e-10 $ \pm $    9.71e-12 \\ 
u & 55799.1311 & 14.20 $ \pm $ 0.04 & 16.49 &  45.18 $ \pm $  1.83 &    2.53e-10 $ \pm $    1.02e-11 \\ 
u & 55799.2087 & 14.20 $ \pm $ 0.04 & 16.54 &  45.20 $ \pm $  1.75 &    2.52e-10 $ \pm $    9.78e-12 \\ 
u & 55799.3988 & 14.00 $ \pm $ 0.04 & 16.35 &  54.60 $ \pm $  2.08 &    3.05e-10 $ \pm $    1.16e-11 \\ 
u & 55799.5373 & 13.88 $ \pm $ 0.04 & 16.28 &  60.71 $ \pm $  2.22 &    3.39e-10 $ \pm $    1.24e-11 \\ 
b & 55797.9261 & 15.66 $ \pm $ 0.06 & 17.70 &  23.92 $ \pm $  1.22 &    1.10e-10 $ \pm $    5.59e-12 \\ 
b & 55797.9930 & 15.50 $ \pm $ 0.05 & 17.61 &  27.78 $ \pm $  1.32 &    1.27e-10 $ \pm $    6.06e-12 \\ 
b & 55799.0032 & 14.18 $ \pm $ 0.04 & 16.58 &  93.67 $ \pm $  3.44 &    4.29e-10 $ \pm $    1.58e-11 \\ 
b & 55799.1318 & 14.12 $ \pm $ 0.04 & 16.53 &  99.37 $ \pm $  3.60 &    4.55e-10 $ \pm $    1.65e-11 \\ 
b & 55799.2094 & 14.04 $ \pm $ 0.04 & 16.50 & 106.18 $ \pm $  3.70 &    4.86e-10 $ \pm $    1.70e-11 \\ 
b & 55799.3995 & 13.88 $ \pm $ 0.04 & 16.33 & 123.39 $ \pm $  4.30 &    5.66e-10 $ \pm $    1.97e-11 \\ 
b & 55799.5382 & 13.79 $ \pm $ 0.04 & 16.27 & 134.55 $ \pm $  4.57 &    6.17e-10 $ \pm $    2.09e-11 \\ 
v & 55797.9309 & 15.27 $ \pm $ 0.06 & 17.18 &  11.12 $ \pm $  0.64 &    4.08e-11 $ \pm $    2.34e-12 \\ 
v & 55797.9978 & 15.10 $ \pm $ 0.06 & 17.09 &  13.12 $ \pm $  0.70 &    4.81e-11 $ \pm $    2.56e-12 \\ 
v & 55799.0058 & 13.93 $ \pm $ 0.05 & 16.14 &  38.48 $ \pm $  1.67 &    1.41e-10 $ \pm $    6.11e-12 \\ 
v & 55799.1360 & 13.79 $ \pm $ 0.04 & 16.11 &  43.45 $ \pm $  1.72 &    1.59e-10 $ \pm $    6.29e-12 \\ 
v & 55799.2135 & 13.75 $ \pm $ 0.04 & 16.09 &  45.33 $ \pm $  1.75 &    1.66e-10 $ \pm $    6.41e-12 \\ 
v & 55799.4037 & 13.65 $ \pm $ 0.04 & 16.00 &  49.86 $ \pm $  1.90 &    1.83e-10 $ \pm $    6.96e-12 \\ 
v & 55799.5432 & 13.49 $ \pm $ 0.04 & 15.90 &  57.34 $ \pm $  2.09 &    2.10e-10 $ \pm $    7.67e-12 \\ 

\enddata
\tablecomments{The full table of photometry is available in the electronic version.}
\end{deluxetable}


\begin{deluxetable}{ccc}
\tablecaption{UV Light Curve Templates}\label{table_templates}
\tablehead {\colhead{Filter} & \colhead{Epoch from Maximum}    & \colhead{Mag}   \\ 
            \colhead{}       & \colhead{(days)} & \colhead{(mag)}  
             } 

\startdata
uvw2 &  -15.0 &  5.875 \\
uvw2 &  -14.9 &  5.793 \\
uvw2 &  -14.8 &  5.712 \\
uvw2 &  -14.7 &  5.632 \\
uvw2 &  -14.6 &  5.553 \\
uvw2 &  -14.5 &  5.474 \\
\enddata
\tablecomments{The epochs and magnitudes are given with respect to the peak time and magnitude in that filter.  The full table is available in the electronic version.}
\end{deluxetable}

\clearpage
\begin{deluxetable}{llrrrrr}
\scriptsize
\tablecaption{Early Count Rate Fits} \label{table_fireballfits}
\tablehead{ \colhead{Filter} &\colhead{Range} & \colhead{$\alpha_1$} & \colhead{t$_0$}& \colhead{$\alpha_2$} & \colhead{t$_{0,2}$}&  \colhead{$\chi^2$/(N-P)\tablenotemark{a}}  \\
\colhead{} & \colhead{days} & \colhead{(erg s$^{-1}$ cm$^{-2}$)} & \colhead{(days)} &  \colhead{(erg s$^{-1}$ cm$^{-2}$)} & \colhead{(days)} & \colhead{()}  
 } 
\startdata
uvw2 & 1-4 &       1.15 $ \pm $     0.08 &       55797.07 $ \pm $     0.09 &  \nodata &  \nodata  &       3.53/ (      11-2) \\
uvm2 &  1-4 &      0.23 $ \pm $     0.048 &       55796.81 $ \pm $      0.28 &   \nodata &  \nodata  &       2.41/ (      11-2) \\
uvw1 & 1-4 &       4.49 $ \pm $      0.30 &       55796.77 $ \pm $     0.09 &  \nodata  &  \nodata  &       11.61/ (      11-2) \\
u & 1-4 &       47.73 $ \pm $       1.48 &      55796.83 $ \pm $     0.04 &  \nodata  &  \nodata  &       7.87/ (      11-2) \\
b & 1-4 &       84.22 $ \pm $       2.58 &       55796.79 $ \pm $     0.03 &   \nodata &  \nodata  &       7.47/ (      10-2) \\
v & 1-4 &       24.57 $ \pm $      0.47 &       55796.62 $ \pm $     0.03 &  \nodata  &  \nodata  &       2.59/ (      11-2) \\
uvw2 & 1-10 &       3.47 $ \pm $      0.39 &       55798.62 $ \pm $      0.17 &   \nodata &   \nodata &       458.80/ (      32-2) \\
uvm2 & 1-10 &       1.93 $ \pm $      0.17 &       55799.76 $ \pm $      0.30 &   \nodata &  \nodata  &       215.24/ (      31-2) \\
uvw1 & 1-10 &       12.09 $ \pm $       1.63 &       55798.08 $ \pm $      0.20 &   \nodata &  \nodata  &       1050.94/ (      29-2) \\
v & 1-10 &       23.65 $ \pm $      0.60 &       55796.57 $ \pm $     0.04 &   \nodata &  \nodata  &       37.93/ (      23-2) \\
uvw2 & 5-10 &       6.29 $ \pm $      0.22 &       55799.71 $ \pm $     0.06 &   \nodata &   \nodata &       7.969/ (      10-2) \\
uvm2 & 5-10 &       2.58 $ \pm $      0.17 &       55800.36 $ \pm $      0.11 &   \nodata &   \nodata &       17.14/ (      9-2) \\
uvw1 & 5-10 &       33.67 $ \pm $       1.23 &       55799.80 $ \pm $     0.06 &   \nodata &   \nodata &       9.88/ (      9-2) \\
v & 5-10 &       31.53 $ \pm $       4.41 &       55797.57 $ \pm $      0.39 &   \nodata &   \nodata &       6.99/ (      4-2) \\
uvw2 & 1-10 &        1.25 $ \pm $      0.06 &        55797.16 $ \pm $      0.07 &        6.87 $ \pm $       0.26 &        55801.23 $ \pm $      0.08 &        12.43/ ( 32-4) \\
uvm2 & 1-10 &       0.26 $ \pm $      0.03 &        55796.96 $ \pm $       0.24 &        3.21 $ \pm $       0.17 &        55801.55 $ \pm $       0.12 &        13.63/ (  31-4) \\
uvw1 & 1-10 &        5.14 $ \pm $       0.27 &        55796.92 $ \pm $      0.07 &        34.49 $ \pm $        1.66 &        55800.95 $ \pm $      0.08 &        38.31/ ( 29-4) \\
v & 1-10 &        23.26 $ \pm $       0.70 &        55796.55 $ \pm $      0.05 &        9.40 $ \pm $        41.23 &        55802.13 $ \pm $        6.56 &        35.19/ (  23-4) \\

\enddata
\tablenotetext{a}{The degrees of freedom are given as the number of points (N) minus 
the number of fit parameters (P)}
\end{deluxetable}


\clearpage 

\end{document}